\documentclass[preprint,review,3p,times]{elsarticle}
\usepackage{amsmath}
\usepackage{graphicx}
\usepackage{lscape} %
\usepackage{comment}
\usepackage{url}%
\usepackage{threeparttable} %
\journal{High Energy Density Physics}

\begin{document}
\bibliographystyle{elsarticle-num}

\begin{frontmatter}
\title{Highly efficient sparse-matrix inversion techniques and 
average procedures applied to collisional-radiative codes}

\author[Saclay]{M. Poirier}%
\ead{michel.poirier@cea.fr}
\author[LIXAM]{F. de Gaufridy de Dortan}

\address[Saclay]{CEA, IRAMIS, Service Photons, Atomes, et Mol\'ecules, 
CE Saclay, b\^at. 522, F91191 Gif/Yvette \textsc{Cedex}, 
\textsc{France}.}
\address[LIXAM]{Laboratoire d'Interaction du rayonnement X avec la 
Mati\`ere, UMR 8624, b\^at. 350, Universit\'e Paris-Sud, F91405 Orsay, 
\textsc{France}.}

\begin{abstract}
The behavior of non-local thermal-equilibrium (NLTE) plasmas plays a 
central role in many fields of modern-day physics, such as laser-produced 
plasmas, astrophysics, inertial or magnetic confinement fusion devices, 
or X-ray sources. The proper description of these media in stationary 
cases requires to solve linear systems of thousands or more rate 
equations. A possible simplification for this arduous numerical task may 
lie in some type of statistical average, such as configuration or 
superconfiguration average. However to assess the validity of this 
procedure and to handle cases where isolated lines play an important 
role, it may be important to deal with detailed levels systems. This 
involves matrices with sometimes billions of elements, which are rather 
sparse but still involve thousands of diagonals. We propose here a 
numerical algorithm based on the LU decomposition for such linear 
systems. This method turns out to be orders of magnitude faster than the 
traditional Gauss elimination. And at variance with alternate methods 
based on conjugate gradients or minimization, no convergence or accuracy 
issues have been faced. Some examples are discussed in connection with 
the krypton and tungsten cases discussed at the last NLTE meeting. 
Furthermore, to assess the validity of configuration average, several 
criteria are discussed. While a criterion based on detailed balance is 
relevant in cases not too far from LTE but insufficient otherwise, 
an alternate criterion based on the use of a fictive configuration 
temperature is proposed and successfully tested. 
It appears that detailed calculations are sometimes necessary, which 
supports the search for an efficient solver as the one proposed here.
\end{abstract}


\end{frontmatter}

\section{Introduction}

Plasmas in a non local thermodynamic equilibrium (NLTE) state play 
an important role in several domains of physics, such as in 
astrophysics, in magnetic and inertial confinement devices, in 
radiation-solid target interaction experiments at large laser 
facilities, or in XUV and X-rays sources\cite{ARa06,Nis06}. 
The proper calculation of important quantities such as the opacity 
or the emission efficiency requires an accurate atomic physics 
description of the active medium. In time-independent regimes 
this implies to solve large systems of kinetic equations 
describing the population transfers between each level. 
Several codes have been developed to deal with NLTE plasmas, among 
which ATOMIC \cite{Hak06}, SCROLL \cite{Bar00}, MOST \cite{Bau04}, 
ATOM3R \cite{Rod06}, AVERROES \cite{Pey01}, FLYCHK \cite{Chu05}, 
CRETIN \cite{Sco01}, NOMAD \cite{Ral01}, JATOM \cite{Sas03}, 
SCAALP \cite{Fau01}. The NLTE meetings provide an opportunity to 
benchmark them \cite{Bow03,Bow06,Rub07,Fon09}.

Because of its inherent complexity, NLTE physics, even in its 
stationary form, lies on efficient linear algebra algorithms. 
In our previous approach to this subject \cite{Poi07}, we relied 
on the standard Gauss algorithm: it proves to be very stable 
and accurate for systems up to several thousand equations, but it  
becomes prohibitively slow above. A possible workaround to this 
issue is to resort to statistical averages, the simplest being the 
configuration average (CA), or the superconfiguration average. 
A discussion of these averaging procedure efficiency may be found 
in the literature (e.g., \cite{Han06}). However, using this average 
one may neglect important properties, for instance when the various 
levels in a given configuration have very different excitation rates. 

This paper investigates two complementary directions to allow the 
treatment of complex NLTE plasmas. First, we discuss various possible 
matrix-inversion algorithms, this step usually being the bottleneck  
of the collisional-radiative codes (Sec.~\ref{sec:matinvtec}).  
Then, in Sec.~\ref{sec:averaging_proc}, we review several 
possibilities aimed at designing a criterion that might efficiently 
qualify the CA procedure. Secs.~\ref{sec:Kr} and \ref{sec:W} 
illustrate this discussion by dealing with two cases considered at 
the last NLTE meeting\cite{Fon09}. Brief concluding remarks are 
given in Sec.~\ref{sec:concl}.

\section{Comparison of sparse-matrix inversion techniques}
\label{sec:matinvtec}

\subsection{System properties}
\begin{figure}[hptb]
\begin{center}
\includegraphics[scale=0.60, angle=0]%
{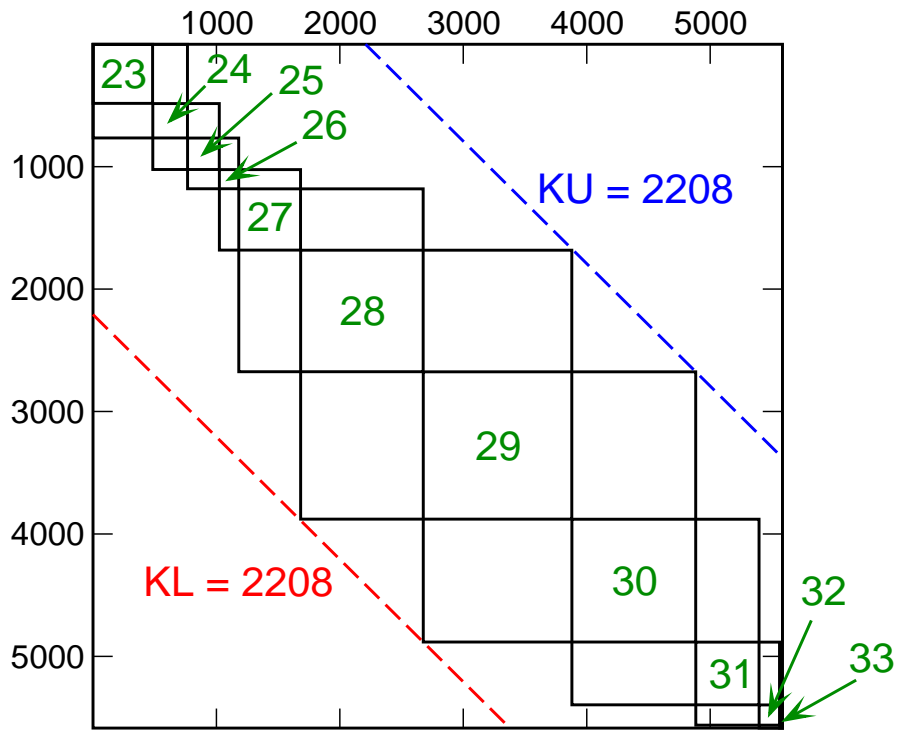}%
\\[+5pt]%
\caption{
Collisional-radiative rate matrix structure: an example in krypton 
at $T_e=1000\text{ eV}$ and $N_e=10^{14}\text{ e/cm}^3$. The number 
of equations is here 5586. The net ion charge is indicated for each 
diagonal bloc. The cumulated number of levels is indicated in 
abscissa and ordinate. The broken lines are the sub- and 
super-diagonal boundaries used in the LU decomposition.
\label{fig:tridiagbloc_kr}}
\end{center}
\end{figure}

In order to properly characterize plasma properties, including 
observables as simple as the average net ion charge, one needs to 
account for a large set of ionic states, including series of 
doubly or even multiply excited states. This implies that the rate 
matrices describing the radiative and collisional transfers in 
such plasmas may have a huge number of elements. 
Because usually the only transition processes considered in 
collisional-radiative systems induce a net charge variation 
$\Delta Z^* = 0,\pm1$, these matrices have a tridiagonal-block 
structure as shown in Fig.~\ref{fig:tridiagbloc_kr}. If a lot of 
$Z^*$ are considered together, and if the numbers of levels 
considered for each ion are similar, the matrix contains a large 
proportion of zeroes and is thus reasonably sparse. Even more, 
inside the tri-diagonal blocs, some transition rates may cancel 
(such as those between multiply and singly excited states). 
However the algorithms proposed here do not account for such zeroes 
inside the tri-diagonal blocs. A direct estimate in a krypton case 
illustrated by Fig.~\ref{fig:tridiagbloc_kr} revealed that 30 to 
40\% of such elements do cancel: for this $5586 \times 5586$ 
matrix, with $3.1\times10^7$ elements, $1.2\times10^7$ are in 
diagonal blocs, which contain themselves $3.8\times10^6$ (32\%) 
``accidentally'' null elements. However, dealing properly with such 
``accidental'' cancelation requires to use a cumbersome indexing 
scheme which may slow down computations, and such attempt was not 
made in the present work.
Clearly this sparse character is stronger when many charge states 
are included and when each bloc has approximately the same size.

\subsection{Gauss algorithm}
The first algorithm used to solve collisional-radiative systems is 
the standard Gauss-Jordan elimination. We could check that this 
method provides in all considered cases a fair numerical accuracy: 
this can be done for instance by canceling the unbalanced transition 
rates and by controlling the level-population departure from the 
Saha-Boltzmann solution \cite{Poi07}. Except when one pivot 
accidentally cancel --- which was never observed in  properly 
defined NLTE cases --- the algorithm always provide a solution,  
through a constant number of operations, which scales as the cube 
of the number of equations. To sum up, the Gauss elimination is 
robust, accurate, but it is really slow and does not take benefit 
from the sparse character of the system. It will be used only when 
we consider small systems --- as those in configuration average 
discussed below --- or when we wish to cross-check other algorithms.

\subsection{Conjugate gradients methods}
Conjugate Gradients (CG) methods are iterative procedures for 
solving a linear system through a series of analytically defined 
steps. The biconjugate gradient method generalize the procedure to 
non symmetric definite positive matrices. The convergence of the 
process may be linked to the use of a preconditioner as discussed, 
e.g., in \cite{Kau90} and will not be considered here. In this 
paper our goal is to check how standard routines widely available 
perform, evaluating their general robustness regardless of the many 
variants that exist. 
One must notice that, at variance with \cite{Kau90} where 
non-stationary systems with few hundreds of levels were included, 
one may deal here with several tens of thousands of equations. 
Using CG methods --- here we use \texttt{linbcg} from 
Numerical Recipes \cite{Pre07} --- one benefits from the sparse 
character of the rate matrix since the user is supposed to provide 
an efficient way to perform the matrix by column vector product.

It appears that the CG codes with no preconditioner may poorly work. 
As seen below, such iterative methods 
start on a user-defined solution, and the convergence may strongly 
depend on this step. Besides a convergence criterion must be chosen. 
It was found that the inline criterion included with \texttt{linbcg} 
routine is not satisfactory: The proposed error parameter may be 
still about 1 while convergence has indeed be reached. Therefore 
in this work we used the natural and physically sound criterion 
based on the average net charge $Z^*$ variation for the last two 
iterations. Namely, when for the first time
\begin{equation}
|<Z^*>_n-<Z^*>_{n-1}| \le \varepsilon \text{\quad and\quad}
|<Z^*>_n-<Z^*>_{n-2}| \le \varepsilon\label{eq:crit_var_Z}
\end{equation}
with for instance $\varepsilon = 10^{-7}$ convergence is assumed 
to be reached. Two steps are considered since it appeared that a 
single check could produce an artificial convergence.

As shown below (subsection \ref{ssec:compar_algorithms}), the CG 
methods when they converge are fast and accurate. However, the CG 
algorithm may lack of robustness. 
For instance, we considered the case where radiative and 
autoionization rates are canceled: including only the collisional 
excitation and ionization and the reverse processes, and comparing 
the solution with Saha-Boltzmann is a way to check the algorithm 
accuracy \cite{Poi07}. In a krypton plasma at a 500 eV temperature 
and $10^{14}\text{ cm}^{-3}$ electronic density, with 5~662 levels 
and 14 charge states accounted for, we did obtain populations 
different from Saha-Boltzmann by $2\times10^{-13}$ in average 
($6\times10^{-10}$ at maximum), but this was obtained after a rather 
large number of iterations: 1966 with $\varepsilon=10^{-13}$ in Eq.
(\ref{eq:crit_var_Z}), starting with zero populations. The inspection 
of $<Z^*>$ before convergence shows strong fluctuations from one 
iteration to the next. 
One may estimate that the condition $\varepsilon=10^{-13}$ is too 
severe, which overestimate the convergence effort, but if one stops 
the process at, e.g., 150 iterations as obtained in well behaved 
cases (see next paragraphs), the $<Z^*>$ fluctuations on the last 
iterations are above $10^{-3}$, which is unacceptably large.

\begin{figure}[tb]%
\begin{center}
	\includegraphics[scale=0.55, angle=0, bb=0 0 612 725, 
	 clip=true]{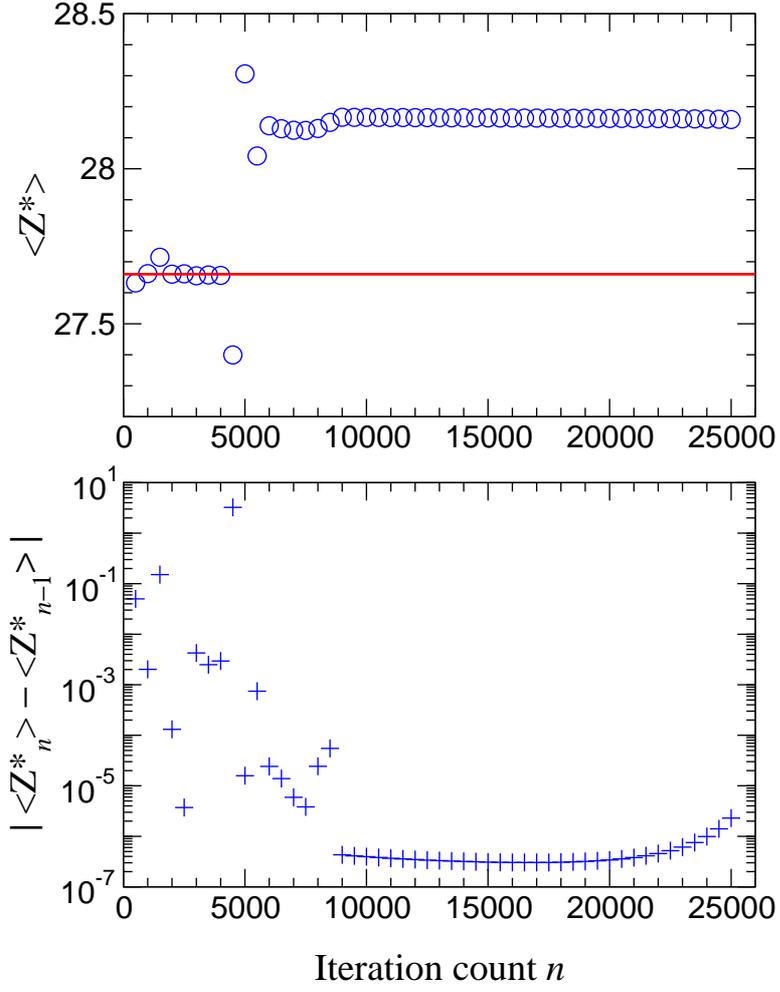}%
	\caption{Average charge (upper figure) and its variation from one 
	step to the previous (lower figure) in Kr as computed with conjugate 
	gradient method starting on Saha-Boltzmann populations. The horizontal 
	red line corresponds to the average charge as determined by the stable 
	Gauss algorithm. The electron temperature and density are 1000 eV and
	$10^{14}\text{ cm}^{-3}$ respectively. The $\varepsilon$ parameter of 
	Eq.~(\ref{eq:crit_var_Z}) is $10^{-13}$. \label{fig:conj_grad_iniSB}}
\end{center}
\end{figure}

\begin{figure}[bt]%
\begin{center}
	\includegraphics[scale=0.55, angle=0, bb=0 0 612 725, 
	 clip=true]{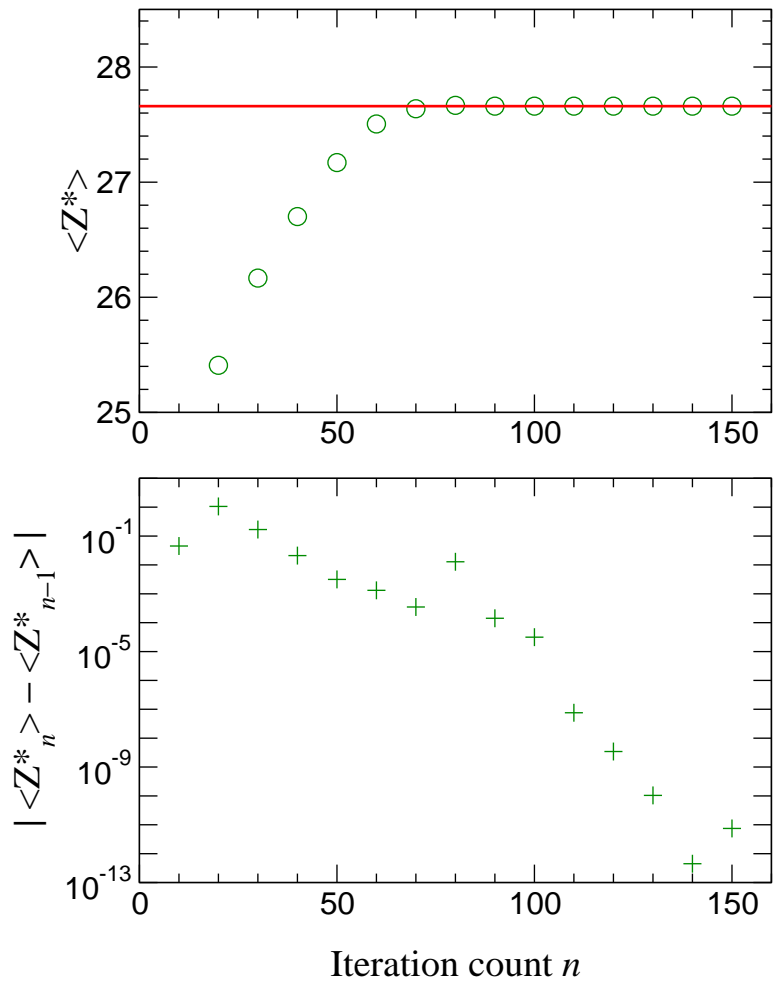}%
	\caption{Same as Fig.~\protect\ref{fig:conj_grad_iniSB} but starting on 
	zero populations. \label{fig:conj_grad_ini0}}
\end{center}
\end{figure}

Moreover, as for every iterative process, the CG convergence may 
depend on the initial choice for the populations. To illustrate 
this when no preconditioner is used, we have plotted in 
Fig.~\ref{fig:conj_grad_iniSB} and Fig.~\ref{fig:conj_grad_ini0}
results of a NLTE analysis in krypton for an electronic temperature 
of $T_e = 1000\text{ eV}$ and an electronic density of $N_e = 
10^{14} \text{ cm}{-3}$. 
Here the number of equations is $N = 5586$, large enough to check 
significantly the calculation speed, but still moderate to allow 
direct computation by the slow but stable Gauss elimination. 
The \texttt{linbcg} internal convergence criterion gives an 
``error'' of 3.3 after 140 iterations and 19.6 after 150 
iterations, while it is obvious from Fig.~\ref{fig:conj_grad_ini0} 
that convergence does then occur. 
One notices that initializing the process with zero 
populations give fast and accurate convergence: 150 steps if 
accuracy of $10^{-12}$ on $<Z^*>$ is sought as indicated by the 
lower part of Fig.~\ref{fig:conj_grad_ini0}; The otherwise known 
correct $<Z^*>$ is given by the horizontal line in the upper figure. 

Conversely, initialization on Saha-Boltzmann populations leads to 
an erratic behavior. After about 4000 steps one gets close to the 
correct solution but with a mediocre $10^{-3}\text{--}10^{-5}$ 
accuracy (cf. lower Fig.~\ref{fig:conj_grad_iniSB}). Then the 
solution begins to diverge and reaches after almost 10~000 
iterations a ``false convergence'' at $<Z^*>\simeq 28.162$, 
significantly off the (otherwise confirmed) Gauss solution of 
$<Z^*>=27.660$. Up to 25~000 iterations have been performed here. 
The $<Z^*>$ variation from one step to the next as seen on the lower 
part of Fig.~\ref{fig:conj_grad_iniSB} remains above $10^{-7}$:  
The fact that this figure is much greater than machine accuracy 
$\simeq10^{-16}$ indicates that this ``convergence'' is not 
satisfactory. This further indicates that the $\varepsilon$ 
parameter of criterion (\ref{eq:crit_var_Z}) must be chosen very 
carefully.

\subsection{LU decomposition}
Another possible choice for matrix inversion is based on the LU 
decomposition algorithm, which uses an upper-lower triangular 
matrix decomposition. The used routine is here \texttt{dgbsvx} 
from \texttt{lapack} linear algebra package, which assumes a 
band-diagonal structure for the matrix to be inverted. The 
balancing of rows and columns is automatically done inside the 
code. The matrix condition is checked and given as an output 
parameter; In the various cases tested, at or far from thermal 
equilibrium, it was found to be satisfactory. In addition to the 
storage of the original matrix with $K_L$ and $K_U$ lower and 
upper diagonals, which occupies $(K_L+K_U+1) \times N$, $N$ being 
the number of equations, the proposed algorithm requires the 
storage of the lower and upper triangular matrices, i.e.,  
$(2K_L+K_U+1) \times N$ additional memory locations. Considering 
realistic values for the $K_L$ ($=K_U$) parameter, this amounts to 
foresee storage as large as 32 GB for a matrix size 
$N \simeq 50~000$.

The various tests performed have shown that this algorithm is fast, 
accurate and robust. No particular row or column balancing had to be 
done by a preprocessor, since this is automatic in the 
\texttt{lapack} routine. The matrix condition is checked inside the 
code and it was found satisfactory in every considered cases. As 
an accuracy test proposed before \cite{Poi07}, we solved with the 
LU algorithm a kinetic system of 4481 equations including only 
collisional excitation, collisional ionization plus the reverse 
processes. One gets then the Saha-Boltzmann solution with a 
maximum population difference of $1.2\times10^{-12}$, and an average 
difference of $5.4\times10^{-16}$, the Gauss algorithm providing  
the same excellent accuracy. It appears that the only serious 
limitation to LU method is the storage issue, even 
for reasonably sparse matrices. Looking again at 
Fig.~\ref{fig:tridiagbloc_kr}, the stored elements are all those 
between the upper and lower broken lines, and additional storage 
is needed for the lower and upper triangular matrices too.

\subsection{Relative efficiency of the algorithms}
\label{ssec:compar_algorithms}
\begin{figure}[tb]%
\begin{center}
	\includegraphics[scale=0.65, angle=0]{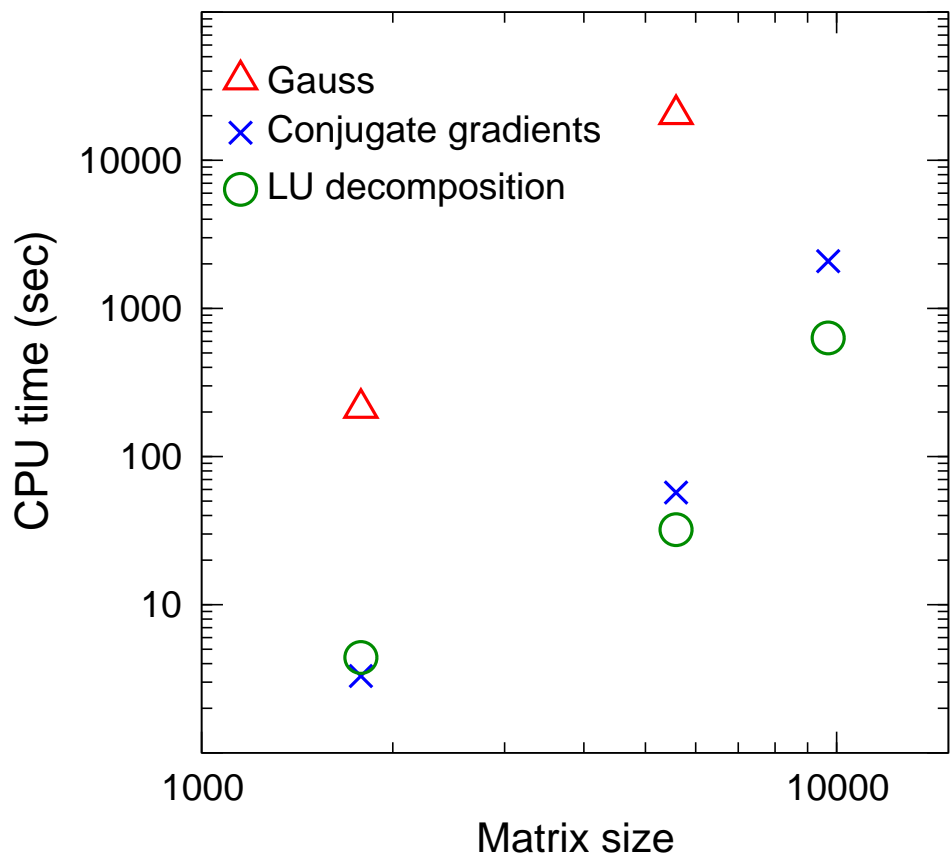}	
	\caption{Computation time for matrix inversion using standard Gauss 
	algorithm, a conjugate gradient code, and LU decomposition.
	\label{fig:inversion_time_Gauss_cg_LU}}
\end{center}
\end{figure}

Examples of elapsed CPU time for the three methods are given in 
Fig.~\ref{fig:inversion_time_Gauss_cg_LU}. Three NLTE cases 
of ``physical interest'' are illustrated here: carbon at 10 eV 
and $10^{12}\text{ e.cm}^{-3}$ with the 7 charge states and about 
1800 levels, krypton at 1 keV and $10^{14}\text{ e.cm}^{-3}$ 
with Kr\textsc{xxiv}--\textsc{xxxiv} ions and about 5600 levels 
included, and at 200 eV, $10^{22}\text{ e.cm}^{-3}$ with 
Kr\textsc{xix}--\textsc{xxviii} ions and about 9700 levels 
included. In the last case, the kinetic system was too large to 
allow ourselves for a resolution by the Gauss algorithm in a 
reasonable time. It appears that while computation time increases 
rapidly with the matrix size, the CG and LU methods are orders of 
magnitude faster then Gauss elimination. Because as seen above the 
tested CG algorithm lacks of robustness, the preferred LU method is 
used in all the computations presented below.

\section{Configuration-average procedures and validity criteria}
\label{sec:averaging_proc}

Configuration average (CA) procedures have been discussed in 
previous papers \cite{Han06, Poi07, Poi08}. It is the simplest and 
more natural way to considerably alleviate the matrix inversion 
task involved in stationary NLTE cases, and, interestingly enough, 
it usually provides rather accurate results.
 
\subsection{Configuration average(s)}
Assuming the ionic level $i$ (resp. $j$) belongs to configuration 
$\alpha$ (resp. $\beta$) the average rate from $\alpha$ to $\beta$ is
\begin{equation}
	R_{\alpha\beta} = \frac1{g_\alpha} \sum_{\substack{i\in\alpha\\
	j\in\beta}} g_iR_{ij}, \text{\quad with\ }
	g_\alpha = \sum_{i\in\alpha}g_i\label{eqn:avratgdef}
\end{equation}
where $g_i$ is the $i$-level degeneracy, and $g_\alpha$ is the 
configuration degeneracy. We can set an alternate definition by 
including the Boltzmann factor
\begin{equation}
	R_{\alpha\beta} = \left.	\sum_{\substack{i\in\alpha\\
	j\in\beta}} g_i \exp(-E_i/(k_BT_e)) R_{ij} \right/
	\sum_{i\in\alpha} g_i \exp\left(-E_i/(k_BT_e)\right),
	\label{eqn:avratgBoltzdef}
\end{equation}
which can be assumed more ``accurate'' because the levels inside 
a given configuration usually have neighboring energies and are 
likely to fulfill thermal equilibrium. Of course, corresponding 
definitions hold for the average energies, energy rms, etc.

Another option that generalizes both previous ones consists in 
arbitrarily defining some ``configuration temperature'', i.e.,
\begin{equation}
	R_{\alpha\beta} = \left.	\sum_{\substack{i\in\alpha\\
	j\in\beta}} g_i \exp(-E_i/(k_BT_c)) R_{ij} \right/
	\sum_{i\in\alpha} g_i \exp\left(-E_i/(k_BT_c)\right),
	\label{eqn:avrat_tconfdef}
\end{equation}
where $T_c$ can be freely chosen. The dependence of the 
various physical quantities on the choice of $T_c/T_e$ 
will qualify the CA validity, as will be illustrated in the next 
sections, in cases where CA holds or does not hold.

\subsection{Validity criteria}
These CA formula being defined, it is important to elaborate a 
criterion that can properly qualify the relevance of this 
procedure, of course avoiding to go back to the time-consuming 
detailed level analysis. 
 
\subsubsection{The natural criterion based energy dispersion}
The most intuitive criterion for the validity of CA may be 
written as
\begin{equation}
	<\Delta E> = \sum_\alpha N_\alpha \Delta E_\alpha \ll k_BT_e
	\label{eqn:crdise}
\end{equation}
where $N_\alpha$ is the $\alpha$-configuration population (with 
$\sum_\alpha N_\alpha=1$) and $\Delta E_\alpha$ the energy 
dispersion
\begin{equation}
	\Delta E_\alpha = \left(\frac{1}{g_\alpha}\sum_{i\in\alpha}
	g_i(E_i-E_\alpha)^2\right)^{1/2}.
\end{equation}
This criterion is fairly easy to implement, since it only requires 
to solve \emph{once} a simple CA kinetic system. However, as 
stressed before \cite{Poi07} and checked once more here (see, e.g., 
subsection \ref{ssec:CA_valid_criter_W}), this necessary condition 
is far from being sufficient, a plain reason being that it involves 
a strongly averaged quantity, which furthermore depends on level 
energies and not on transition rates that can vary significantly 
inside a given pair of configurations.

\subsubsection{A criterion based on LTE test and its applicability 
range}\label{ssec:CA_valid_LTE}
A criterion based on the comparison between the Saha-Boltzmann 
solution and of the solution of a \emph{partial rate} kinetic 
equation has been proposed in previous papers \cite{Poi07,Poi08}. 
It consists in checking how well the CA of a system containing 
rates that obey \emph{detailed balance} --- 
collisional rates in \cite{Poi07}, any rates in \cite{Poi08} ---
reproduces the Saha-Boltzmann solution. In more detail, one first 
evaluates using (\ref{eqn:avratgdef}) average collisional rates 
$R'_{\alpha\beta}$ where the rates $R'$ include \emph{only the 
collisional excitation, the collisional ionization, and the 
reverse processes}. Then one solves the CA kinetic system
\begin{equation}
	\frac{dN_\alpha}{dt} = -\sum_{\substack{\beta\\\beta\ne\alpha}} 
	R'_{\alpha\beta} N_\alpha +\sum_{\substack{\beta\\\beta\ne\alpha}} 
	R'_{\beta\alpha} N_\beta = 0	\label{eqn:avrateq}
\end{equation}
that gives the configuration populations $N_\alpha$. Last, 
one compares this solution with the Saha-Boltzmann populations of
the configurations $N^\text{SB}_\alpha$. Since the $R'$ rates obey 
detailed balance, the difference $N_\alpha-N^\text{SB}_\alpha$ is 
a measure of the validity of CA. This criterion has proven to give 
good results in a carbon plasma at 3 eV or 10 eV \cite{Poi07}. 

However we must point some drawbacks of this approach. First, the 
criterion in its simplest form relies on a comparison of average 
net charges, which is known to be rather insensitive to the detail 
of the populations, regarding excited levels for instance. Second 
and more seriously, this criterion checks the CA validity close to 
LTE, and, if the studied situation is \emph{significantly far from 
LTE}, it cannot be predictive. Examples in this direction will be 
provided in next sections. This criterion appears as necessary but 
not sufficient either.

\subsubsection{An alternate criterion based on fictive 
configuration temperatures}\label{ssec:config_temp}
A last approach consists in solving the CA kinetic equations 
using the various averages [Eqs. (\ref{eqn:avratgdef}), 
(\ref{eqn:avratgBoltzdef}), (\ref{eqn:avrat_tconfdef})] and to 
compare the results. The comparison may concern several physical 
quantities such as $<Z^*>$, the radiative bound-bound rate 
\begin{equation}
P_{\text{bb}}	= N_\text{ions} \sum_i p_i \sum_j (E_i-E_j)A_{ij}
\end{equation}
$p_i$ being the $i$-level population, $E_i-E_j$ the 
transition energy, and $A_{ij}$ the radiative rate, or the 
radiative bound-free rate. This criterion requires to solve 
at least two CA systems, and preferably three, one at $T_c$ 
infinite, one at $T_c=T_e$, and one at a fictive low $T_c$, 
e.g., $0.1T_e$. Examples will be given in the next sections.

\section{The krypton case: validity and limitation of CA}
\label{sec:Kr}

The analysis of charge distribution and radiative losses in krypton 
presents a definite interest for various reasons. Krypton is used in 
numerous plasmas devices, and one must mention the availability of 
radiative loss measurements by Fournier \emph{et al} \cite{Fou00}.

The present study deals with temperatures at least equal to 
500 eV, and electron densities up to $10^{22}\text{ cm}^{-3}$, 
charge states from XIX to XXXVII are accounted for. For such large 
temperatures, the collisional rates are low enough to provide a 
strongly non-LTE situation. The configurations included in the 
computation are listed in Table \ref{tab:config_Kr}. In order to 
include an important set of excited states required for a 
pertinent description, this list is much more complete than the 
one considered for test purposes in Sec. \ref{sec:matinvtec}. For 
instance at 1 keV, $10^{14}\text{ e/cm}^3$, the present computation 
includes 45~927 levels vs 5~587 previously.  
The energy and rate computations in krypton as well as in tungsten 
are performed using the HULLAC suite \cite{Bar01}.

This accounts for a large number of autoionizing configurations while 
keeping the NLTE computation tractable even in the detailed case. Lower 
temperatures would require a more complete set of ions but are within 
the reach of present method. 

\begin{figure}[tb]%
\begin{center}
	\includegraphics[scale=0.625, angle=0]{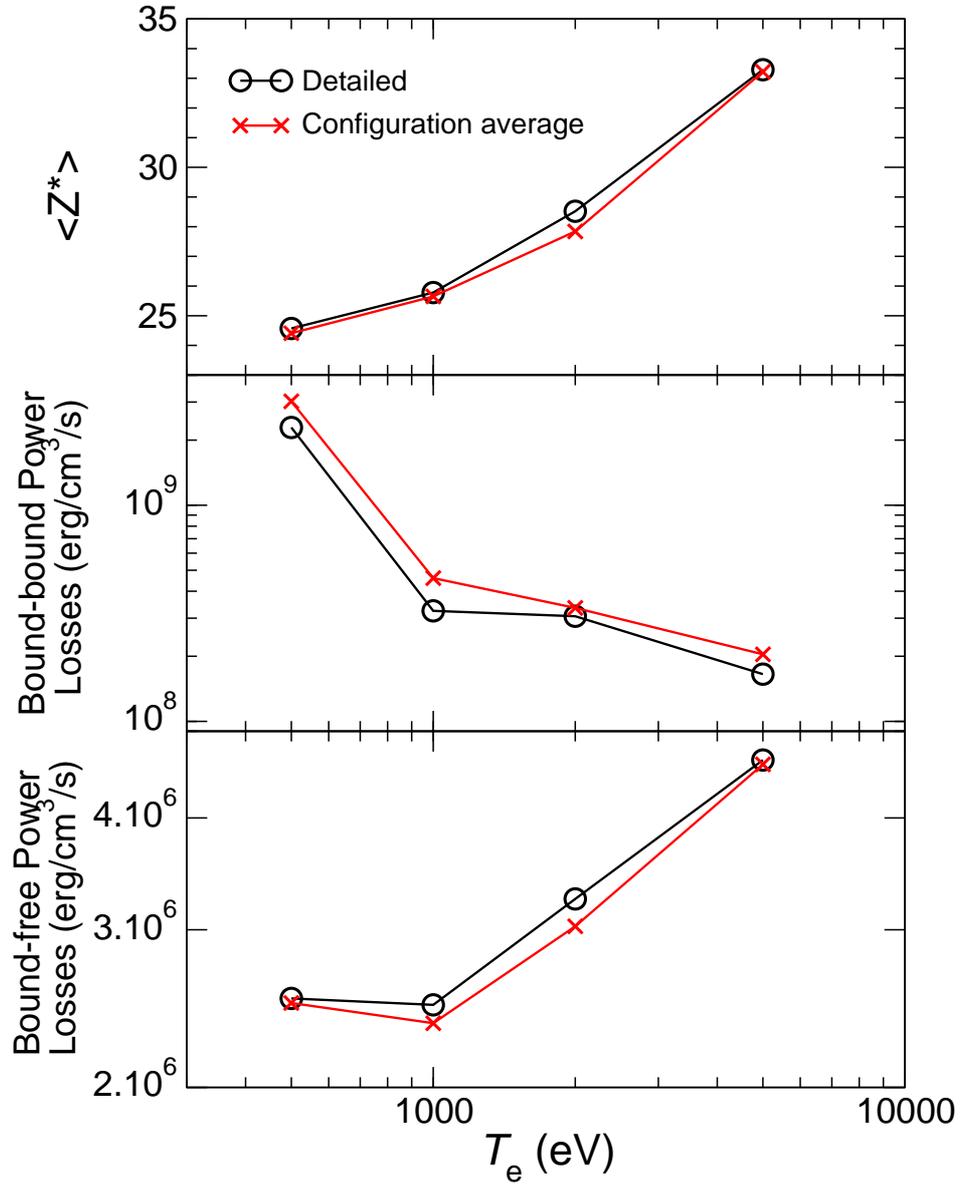}	
	\caption{Average charge, radiative bound-bound and bound-free losses
	for a krypton plasma at $10^{14}\text{ cm}^{-3}$ 
	electronic density as a function of the temperature in eV. Computations 
	are performed using detailed levels and configuration average.
	\label{fig:Kr_Ne14_Zbar_ploss}}
\end{center}
\end{figure}

\begin{figure}[tb]%
\begin{center}
	\includegraphics[scale=0.625, angle=0]{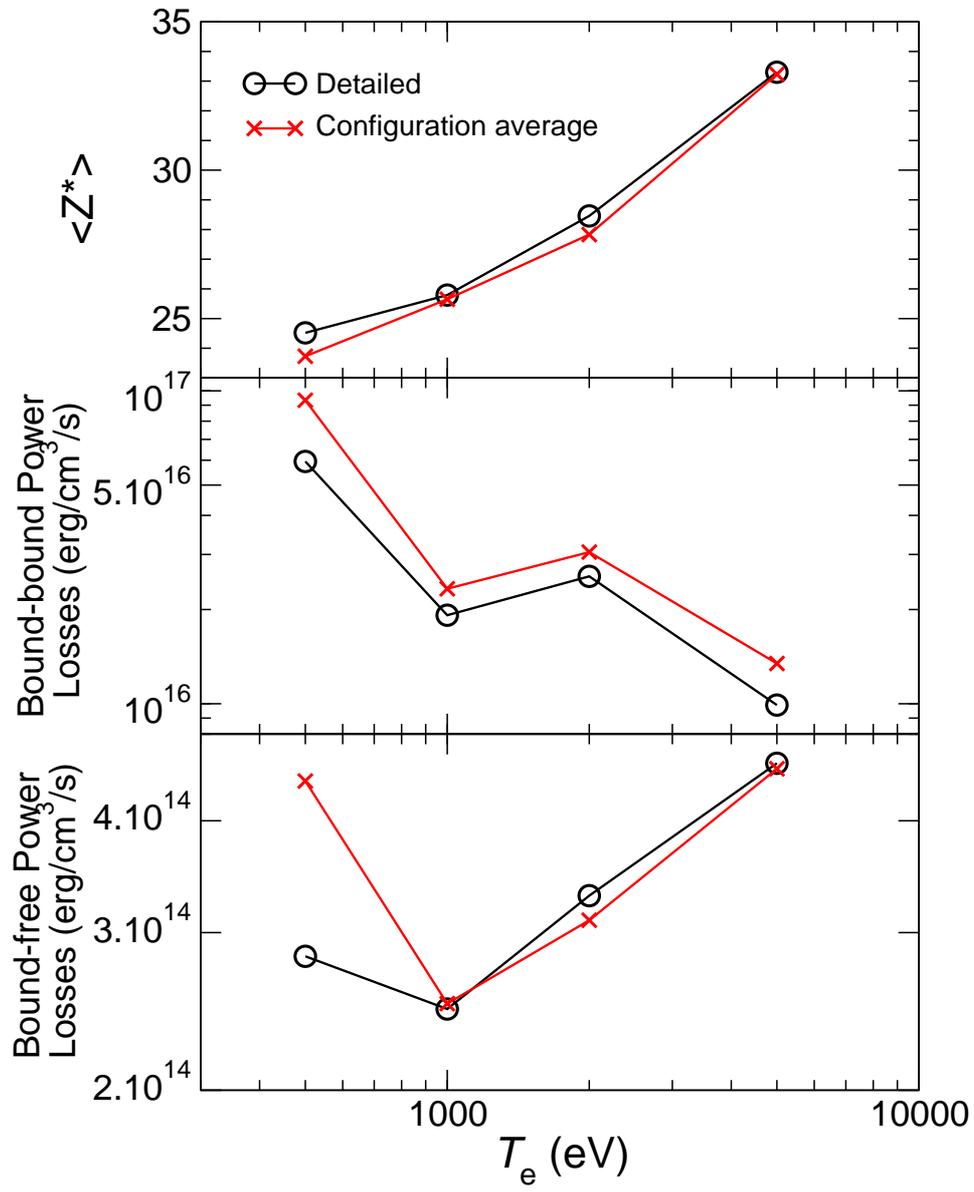}	
	\caption{Same as Fig.~\protect\ref{fig:Kr_Ne14_Zbar_ploss} but for an 
  electronic density of $10^{18}\text{ cm}^{-3}$.
  \label{fig:Kr_Ne18_Zbar_ploss}}
\end{center}
\end{figure}

The average charge, the radiative bound-bound (bb) losses and the 
radiative bound-free (bf) losses are plotted in 
Fig. \ref{fig:Kr_Ne14_Zbar_ploss} and \ref{fig:Kr_Ne18_Zbar_ploss} 
for electronic densities equal to $10^{14}$ and 
$10^{18}\text{ cm}^{-3}$ respectively. Detailed computation and 
configuration average results are both displayed. Once again, 
the agreement between both computations is good. This is particularly 
true at $T=1000\text{ eV}$ and $T=5000\text{ eV}$ where the average 
charge is close to, respectively, 26 and 34, which corresponds to 
the Ne-like and He-like closed shell ions. Then it is well known that 
all computations tend to give agreeing predictions. 

Since Fournier \emph{et al} performed measurements of the radiative 
cooling coefficient in coronal Kr plasmas \cite{Fou00}, it is 
instructive to compare our $10^{14}\text{ cm}^{-3}$ data to theirs. 
The radiative cooling coefficient is defined for coronal plasmas as
\begin{equation}
L^\text{tot} = <Z^*>(P_\text{bb}+P_\text{bf}+P_\text{ff})/N_e^2
\end{equation}
where $P_\text{ff}$ is the Bremsstrahlung loss rate, which we can 
derive from the approximate expression (formula 4.24.4 in 
\cite{Wes04})
\begin{equation}
	P_\text{ff} \text{(erg/s/cm$^3$)} \approx 5.35\times10^{-24}
	<Z^*> N_e^2 (T_e)^{1/2}
	\text{ with $N_e$ in cm$^{-3}$, $T_e$ in keV.}\label{eq:ff}
\end{equation}
More accurately, we should substitute $<Z^*{}^2>/<Z^*>$ to $<Z^*>$ 
in the rate (\ref{eq:ff}) but this looks unnecessary since this is 
just an estimation. At 1 keV and $10^{14}\text{ e.cm}^{-3}$, the 
ff term is $P_\text{ff}\simeq1.4\times10^6\text{ cgs}$, less than 
the bf one and quite small with respect to the bb contribution.

From Fig. 6 of Ref.\cite{Fou00}, we get a radiative coefficient 
$L^\text{tot} = 6\times10^{-32}\text{ W.m}^3$ at 1 keV, 
while the present determination is $7.8\times10^{-32}$, in very good 
agreement if we consider the difficulty of this measurement. 
At 500 eV, the agreement deteriorates ($1.6\times10^{-31}\text{ SI}$ 
measured, vs. $5.9\times10^{-31}$ computed), while at 2 keV one gets 
again fair agreement ($6\times10^{-32}$ measured vs 
$8.6\times10^{-32}$ obtained here).

A clear discrepancy between detailed and configuration average is 
visible on the bound-free losses at 500 eV and $10^{18}\text{ 
cm}^{-3}$: $2.82\times10^{14}\text{ erg/s/cm}^3$ in the detailed 
model, $4.43\times10^{14}\text{ erg/s/cm}^3$ in CA. Nevertheless 
this bf-loss ratio does not exceed 1.57, and bound-free losses are 
two orders of magnitude lower than bound-bound losses, which means 
that such a difference will hardly be noticed experimentally, only 
the sum being measured. 
However, one may try to look closer at the origin of this discrepancy. 
A first explanation could come from the averaging procedure that 
can include or not the Boltzmann factor (Eqs. \ref{eqn:avratgBoltzdef}, 
\ref{eqn:avratgdef}), the Fig.~\ref{fig:Kr_Ne18_Zbar_ploss} 
corresponding to the case including the Boltzmann factor. The 
inspection of Table \ref{tab:Kr_Ne18_gBoltz_vs_g} demonstrates that 
the averaging procedure is not at stake. For all the plotted data, 
both configuration-average quantities would be indistinguishable at 
the drawing accuracy. One simply notices that the agreement between 
both averages increases with $T_e$, as expected.

A more thorough analysis arises from the detailed analysis of 
the radiative losses contributions. The configurations contributing 
the most to bound-free losses according to the CA calculations are 
listed in Table \ref{tab:Kr_Ne18_dominant_contrib_bf}. 
One may see that, e.g., the $1s^22s^22p^63p3d$ configuration of 
$\text{Kr}^{24+}$ dominates for the bf losses in the CA computation
(total population 0.0147, contribution to bf losses 
$2.48\times10^{14}\text{ erg/s/cm}^3$), while the corresponding 
levels contribute for a population of 0.0219 and bf losses of 
$5.13\times10^{13}\text{ erg/s/cm}^3$ in the detailed computation. 

Conversely, in the \emph{detailed} scheme, the level contributing 
the most to bf losses is the Kr$^{25+}$ ground level 
($8.46\times10^{13}\text{ erg/s/cm}^3$), followed by the 
Kr$^{24+}$ $(1s^22s^22p^6)^1S_0 3p_{1/2}3d_{5/2} J=3$ level 
($4.68\times10^{13}$) and Kr$^{26+}$ Ne-like ground level 
($4.65\times10^{13}$). 

As a rule one notices that a detailed inspection (e.g., at the 
configuration level) of populations or radiative losses may reveal 
large discrepancies between models that will remain hidden when only 
global quantities are considered.

To check how the CA validity criteria defined in subsection 
\ref{ssec:config_temp} behaves, we have done two additional 
computations of the Kr plasma properties in configuration average. 
One was performed using the ``infinite configuration temperature'' 
scheme as defined by formula (\ref{eqn:avratgdef}), while the last 
one used a finite temperature (\ref{eqn:avrat_tconfdef}) chosen 
below the electronic temperature $T_c/T_e=0.1$. The results of the 
various CA compared to the detailed computation are shown in Table
\ref{tab:Kr_Ne18_gBoltz_vs_g}. It appears that the average 
(\ref{eqn:avrat_tconfdef}) agrees usually within 2\% or better 
with (\ref{eqn:avratgdef}), except on bf losses at 500 eV, where 
the difference exceeds 5\%. This discrepancy is an indication of 
the (stronger) difference between the CA and detailed values. A 
computation with $T_c/T_e=0.01$ would increase this difference.

\section{The tungsten case: an example of configuration-average 
validity breakdown}\label{sec:W}

Tungsten exhibits some interesting properties that qualify it for 
being one of the constituents of the divertor in modern tokamaks 
\cite{Hir07}. It presents a high melting point, a low erosion 
rate, and a low tritium retention. However, in such a plasma one 
may expect high radiative losses because of the richness of its 
spectrum. It also allows to benchmark X-ray spectrum calculations
\cite{Ral06}. That is why tungsten was in the list of cases 
submitted to the NLTE-5 workshop \cite{Fon09}.

\subsection{Description of the computation}
The list of configurations used in the present computation is 
similar to the one used in krypton (Table \ref{tab:config_Kr}). The 
computations performed here are restricted to $T_e = 20\text{ keV}$ 
and 30 keV, at a density of $N_e=10^{14}\text{ cm}^{-3}$. The 
density $10^{24}\text{ cm}^{-3}$ was also considered, but since 
the present model does not account for continuum lowering (or 
pressure ionization), these results are not really significant and 
could be exploited only if compared to alternate models. The net 
charge states included are 59--72 and 61--74 for $T_e = 20
\text{ keV}$ and 30 keV respectively (at $10^{24}\text{ cm}^{-3}$, 
60--74 and 61--74 charges were considered). 

\subsection{Average charge and radiative losses analysis}
Results are summarized in Table \ref{tab:W_Ne14_gBoltz_vs_g}. 
Considering the $<Z^*>$ value, we check that detailed and 
configuration-average computations provide similar figures: 
at 20 keV one gets 65.004 and 64.547 respectively, and at 
30 keV, 67.389 and 67.265, where the CA is performed using 
Boltzmann ponderation as defined in (\ref{eqn:avratgBoltzdef}). 
The bound-free radiative losses, mostly sensitive to the 
ground-state population of each ions, are also quite similar in 
both computations. 
However, a huge discrepancy is observed on the bound-bound 
radiative losses, since at 30~000 eV the computed figures are 
$4.866\times10^7$ and $6.496\times10^{13}$ erg/s/cm$^3$ in the 
detailed and CA cases respectively. 

\begin{figure}[tb]%
\begin{center}
	\includegraphics[scale=0.65, angle=0]{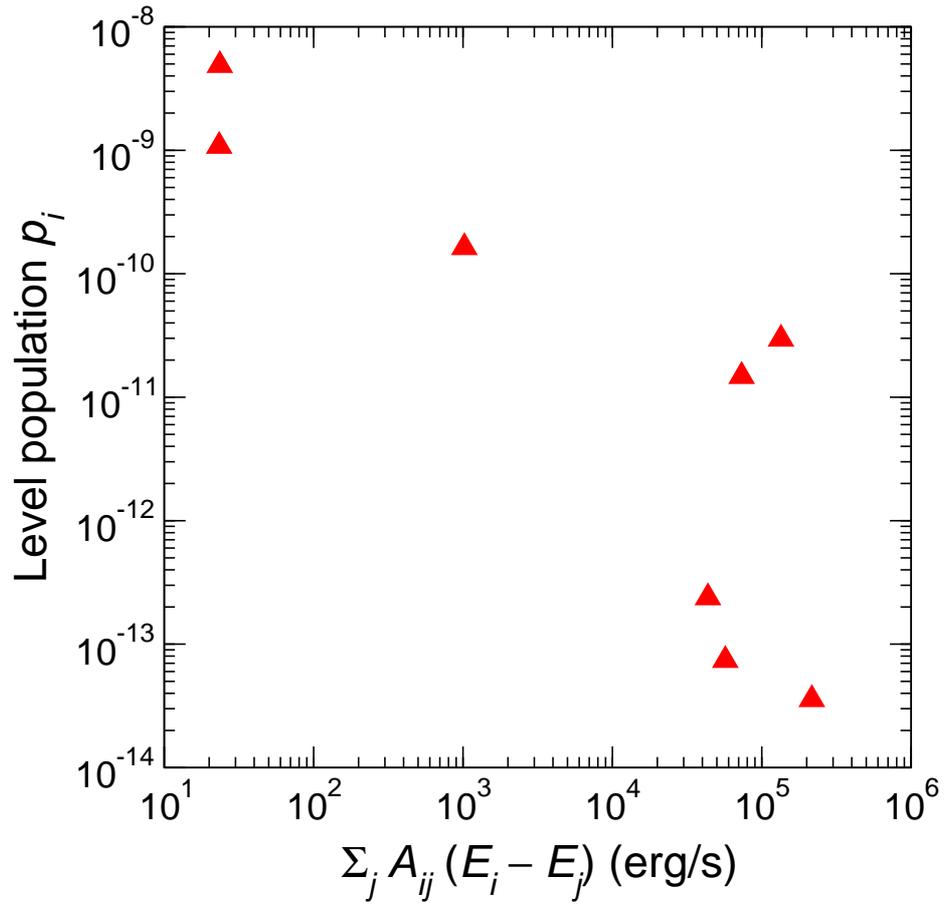}	
	\caption{Radiative losses from and populations of the eight levels 
	belonging to the $1s^22s2p^2$ configuration of $W^{69+}$. The 
	lower levels $j$ of the radiative decay belong to the $1s^22s^22p$
	configuration. The detailed level computation is done at 30~000 eV 
	and $10^{14}\text{ e/cm}^3$. \label{fig:2s2p2_popi_vs_AEij}}
\end{center}
\end{figure}

This surprising feature cannot be explained by a drastic variation 
of the charge distribution, rather similar in both cases. 
To trace the origin of this discrepancy, one must resort to a 
finer analysis of the contributions of each pair of configurations 
to the bound-bound rate. Doing this, it appears that the huge 
bound-bound losses in CA case originate mostly from the transition 
between the $1s^22s^22p$ and the $1s^22s^22p$ configurations of the 
W\textsc{lxx} ion. While the former configuration exhibits a plain 
fine-structure splitting of the $2p$ doublet (of about 1500 eV, 
which much less than the electronic temperature), the latter one 
has a more complex structure. It contains 8 levels, the 5 lowest 
ones being at energies close to one of the $1s^22s^22p$ doublet 
component, while the 3 upper levels are approximately 1700 eV 
above the $2p_{3/2}$ upper component of the doublet. The two 
configuration levels therefore strongly overlap. Considering the 
radiative deexcitation rates, a detailed inspection demonstrates 
that the 5 higher levels of the $1s^22s^22p$ configuration exhibit 
the largest radiative rates (above $10^{13}\text{ s}^{-1}$), while 
they are noticeably the less populated. This fact 
is illustrated by Fig. \ref{fig:2s2p2_popi_vs_AEij}, where the 
eight level populations $p_i, i\in 1s^22s2p^2$ as obtained from 
the detailed computation are plotted in ordinate while the 
abscissas stand for the individual radiative loss factor 
$\sum_{j\in 1s^22s^22p} A_{ij}(E_i-E_j)$. The anticorrelation 
between populations and rates appear strikingly on this plot. An 
additional explanation for the inadequacy of the CA computation in 
this case comes from the \emph{total population} of the $1s^22s2p^2$ 
configuration. The CA computation gives $\simeq4\times10^{-3}$, 
while the sum of the eight level populations in the detailed 
approach is only $\simeq3\times10^{-8}$, lowering thus the 
radiative bound-bound losses by 5 orders of magnitude. The 
discrepancy on this configuration population originates also from 
significant variations in the collisional and radiative detailed 
rates between configurations, and it has not been analyzed in more 
detail.

\subsection{About various criteria assessing the validity of 
configuration average}\label{ssec:CA_valid_criter_W}
One must now consider whether this strong discrepancy was expected, 
since the CA average remain much less computer-time consuming.

First the plain criterion on the average energy dispersion is 
pointless here: one computes for the energy rms inside each 
configuration ponderated by its population (\ref{eqn:crdise})
$<\Delta E> = 625\text{ eV}$, far below $T_e = 30~000\text{ eV}$. 
Then, the criterion of subsection \ref{ssec:CA_valid_LTE} based on 
the comparison of the Saha-Boltzmann solution with the CA solution 
of a modified rate equation is of no help either. The temperature 
being very high (and the density very low), the modified rate 
equations give a plain fully ionized plasma, the $Z^* < 74$ ion 
population being less than $10^{-7}$. This solution is in excellent 
agreement with Saha-Boltzmann equation, but provides no relevant 
information on the real plasma, far from thermal equilibrium. 

One must then resort to the fictive-configuration-temperature 
criterion devised in subsection \ref{ssec:config_temp}. The use of 
this criterion is illustrated in Table \ref{tab:W_Ne14_gBoltz_vs_g}. 
Therefore, in addition to the detailed computation, we have 
performed the three configuration averages proposed in Sec. 
\ref{sec:averaging_proc}. The first average (\ref{eqn:avratgdef}) 
is the $g$ ponderation, i.e., the infinite temperature limit; the 
``natural'' average (\ref{eqn:avratgBoltzdef}) involving Boltzmann 
factors corresponds to the previously discussed CA value. 
For the ``fictive configuration temperature'' case 
(\ref{eqn:avrat_tconfdef}) one has taken $T_c/T_e=0.1$, in order 
to test $T_c$ well above and below $T_e$. 

Considering first the $Z^*$ average, comparing the CA data obtained 
with an infinite temperature (column 3) with the data with low 
$T_c$ (column 5, the column 4 being in between), one checks that 
these two figures are close. If we look now at the b-b losses 
(columns 6--9), we see that (\ref{eqn:avratgdef}), or 
(\ref{eqn:avratgBoltzdef}) which is close, differ by almost a  
factor of 2 at 20~000 eV, and less but still significantly at 
30~000 eV. \emph{This discrepancy between averages 
(\ref{eqn:avratgdef}) and (\ref{eqn:avrat_tconfdef}) must be 
interpreted as a failure of the configuration average}, i.e., is 
a clue of the even much stronger discrepancy between the detailed 
result and the (\ref{eqn:avratgBoltzdef}) average. Once again, the 
interesting fact in using (\ref{eqn:avratgdef}) and 
(\ref{eqn:avrat_tconfdef}) is that this requires only two 
quite fast \emph{configuration average} computations to check 
whether a detailed analysis is required. Moreover, looking at 
$Z^*$, bound-bound and bound-free losses in this table, one 
notices immediately that only the bound-bound figure in CA 
formalism is highly dubious.

In order to validate further the $T_c$ criterion, we also performed 
computations with $T_c/T_e=0.01$. It appears that, at 20~000 eV, 
one gets $<Z^*>=64.362$ with $T_c=200\text{ eV}$, not too different 
from the $T_c \infty$ limit, which strengthen our confidence in 
this CA average. Conversely, in the same conditions, the bound-bound 
losses are $2.2\times10^9\text{ erg/s/cm}^3$ to be compared to the 
$T_c$-infinite CA value of $1.3\times10^{13}\text{ erg/s/cm}^3$: 
This huge change clearly indicates a failure of the CA 
approximation when dealing with bb losses. Once again, the 
modest variation of the bound-free losses ($8.48\times10^6
\text{ erg/s/cm}^3$ to $8.60\times10^6\text{ erg/s/cm}^3$ when 
$T_c$ varies from $T_e/100$ to $\infty$) insures that such 
observable is reasonably computed within CA approximation.

\section{Conclusions}\label{sec:concl}

Reliable computations of plasma properties significantly out of 
local thermal equilibrium involving complex atoms require to solve 
large linear systems, for which various algorithms have been tested.
Without claiming exhaustivity, it appears that the LU decomposition 
method offers a good compromise regarding stability, robustness, 
accuracy and computation speed. Alternate methods using, e.g., 
conjugate gradients with a preconditioner are worth consideration 
but certainly of less direct usage. We have investigated a second 
direction to simplify the kinetic system solution, based on 
configuration average. In addition to the already known criteria, 
that provide a necessary but not sufficient condition, we have 
tested a criterion based on the use of a fictive configuration 
temperature. This criterion proves to be very efficient including 
when dealing with situations far off thermal equilibrium. Moreover, 
even negative configuration temperatures may be used, allowing a 
vast range of configuration average checking. Since in some 
situations, as illustrated by the bound-bound power losses in 
tungsten at high $T_e$, this average performs poorly it is 
essential to use an efficient validity criterion and when necessary 
to dispose of a fast and robust algorithm for the very large kinetic 
systems that must then be solved.

\section*{Acknowledgements}
The authors gratefully acknowledge stimulating discussions with 
T. Blenski and the decisive assistance from E. Audit on 
computational issues. 
They also thank A. Bar-Shalom, M. Busquet, M. Klapisch, 
and J. Oreg for making the HULLAC code available. This work has 
been partly supported by the European Communities under the 
contract of Association between EURATOM and CEA within the 
framework of the European Fusion Program. The views and opinions 
expressed herein do not necessarily reflect those of the European 
Commission.

\bibliography{jrnlabbr,colrad3}

\clearpage
\section*{Tables}

\begin{threeparttable}[thb] %
	\centering\footnotesize%
		\begin{tabular}{c c c p{9cm}}
		\hline\hline
		Ion & $N_c$ & $N_l$ & List of configurations \\
		\hline
KrXIX & 70 & 3781 & $[1s^2 2s^2 2p^6] 3s^2 3p^5 nl$, $3s^2 3p^4 3d nl$, 
$3s 3p^6 nl$, $3p^6 3d nl$; 
 $1s^2 2s^2 2p^5 3s^2 3p^6 nl$, $1s^2 2s 2p^6 3s^2 3p^6 nl$ \\			
\vtop{\hbox{KrXX--XXII}\vspace{1ex}\hbox{$(N=4,3,2)$}} & 63 
& {}\tnote{a} & 
$[1s^2 2s^2 2p^6] 3s^2 3p^N nl$, 
$3s^2 3p^{N-1} 3d nl$, $3s 3p^{N+1} nl$; 
  $1s^2 2s^2 2p^5 3s^2 3p^N nl$, $1s^2 2s 2p^6 3s^2 3p^N nl$ \\
KrXXIII & 80 & 6006 & $[1s^2 2s^2 2p^6] 3s^2 3p nl$, $3s^2 3d nl$,  
$3s 3p^2 nl$, $3s 3p 3d nl$, $3p^3 nl$, $3p^2 3d nl$;  
$1s^2 2s^2 2p^5 3s^2 3p^2 nl$, $1s^2 2s 2p^6 3s^2 3p^2 nl$ \\
KrXXIV & 75 & 2239 & 
$[1s^2 2s^2 2p^6] 3s^2 n$l, $3s 3p nl$, $3s 3d nl$, 
$3p 3d nl$; 
  $1s^2 2s^2 2p^5 3s^2 3p nl$, $1s^2 2s 2p^6 3s^2 3p nl$ \\
KrXXV & 61 & 1499 & $[1s^2 2s^2 2p^6] 3s nl$, $3p nl$, $4l 4l'$,
$1s^2 2s^2 2p^5 3l 3l' 3l''$; %
  $1s 2s^2 2p^6 3l 3l^\prime 3l^{\prime\prime}$ \\
KrXXVI & 67 & 2129 & $1s^2 2s^2 2p^6 nl$, $1s^2 2s^2 2p^5 nl n^\prime 
l^\prime$, $1s^2 2s 2p^6 3l 3l^\prime$;
  $1s 2s^2 2p^6 3l 3l'$ \\
KrXXVII & 83 & 5397 & $1s^2 2s^2 2p^5 n$l, $1s^2 2s 2p^6 nl$, 
$1s^2 2s^2 2p^4 n l n' l'$, 
$1s^2 2s 2p^5 3l 3l^\prime$, $1s^2 2p^6 3l 3l^\prime$;
  $1s 2s^2 2p^6 3l$ \\
KrXXVIII & 82 & 6758 & $1s^2 2s^2 2p^4 nl$, $1s^2 2s 2p^5 nl$, 
$1s^2 2s^2 2p^3 n l n' l'$; 
  $1s 2s^2 2p^6$ \\
KrXXIX & 74 & 4739 & $1s^2 2s^2 2p^3 nl$, $1s^2 2s 2p^4 nl$, 
$1s^2 2p^5 nl$, $1s^2 2s^2 2p^2 nl n'l'$; 
  $1s 2s^2 2p^5$, $1s 2s 2p^6$ \\
\vtop{\hbox{KrXXX--XXXI}\vspace{1ex}\hbox{$(N=2,1)$}} & 94 & 
{}\tnote{b} & $1s^2 2s^2 2p^N nl$, $1s^2 2s 2p^{N+1} nl$, 
$1s^2 2p^{N+2} nl$, $1s^2 2s^2 2p^{N-1} nl n'l'$; 
  $1s 2s^2 2p^{N+2}$, $1s 2s 2p^{N+3}$, $1s 2p^{N+4}$, 
  $1s 2s^2 2p^{N+2} 3l$, $1s 2s 2p^{N+2} 3l$, $1s 2p^{N+3} 3l$ \\
KrXXXII & 135 & 3568 & $1s^2 2s^2 nl$, $1s^2 2s 2p nl$, 
$1s^2 2p^2 nl$, $1s^2 2s n l n' l'$; 
  $1s 2s^2 2p^2$, $1s 2s 2p^3$, $1s 2p^4$, $1s 2s^2 2p 3l$, 
  $1s 2s 2p^2 3l$, $1s 2p^3 3l$ \\
KrXXXIII & 80 & 751 & $1s^2 2s nl$, $1s^2 2p nl$, $1s^2 nl n'l'$; 
  $1s 2s^2 2p$, $1s 2s 2p^2$, $1s 2p^3$, $1s 2s^2 3l$, 
  $1s 2s 2p 3l$, $1s 2p^2 3l$ \\
KrXXXIV & 23 & 106 & $1s^2 nl$; $1s 2s^2$, $1s 2p^2$, $1s2l3l'$ \\
KrXXXV & 18 & 59 & $1s nl$; $2l 2l'$ \\
KrXXXVI & 15 & 25 & $nl$ \\
\hline\hline
		\end{tabular}
\begin{tablenotes}
\item[a] $N_l = 6988, 8980, 7319\text{ for }N=4,3,2$ respectively
\item[b] $N_l = 3607, 1915\text{ for }N=2,1$ respectively
\end{tablenotes}
	\caption{Configurations included in Kr computation. Unless 
	otherwise specified, one has $n\le 5, l\le n-1$. When $n$ and $n'$ 
	are involved in the same configuration, one has $6\le n+n'\le 8$
	with $l\le n-1, l'\le n'-1$. $N_c$ is the number  
	of nonrelativistic configurations, $N_l$ the number of levels. In the 
	HULLAC computation, the configurations are usually separated in two 
	groups within which interaction is fully accounted for; these groups 
	are separated by a semicolon in the list, the second group containing 
	excitation of an inner electron ($1s$, $2s$ or $2p$).}
	\label{tab:config_Kr}
\end{threeparttable}

\renewcommand{\arraystretch}{1.0}
\begin{table}[hbtp]
\centering%
\begin{tabular}{cccccc}
\hline\hline
 $Z$ & Configuration & CA: population & CA: bf losses & 
  DL: population &  DL: bf losses \\[-3pt]
  &&&(erg/s/cm$^3$)&&(erg/s/cm$^3$)\\
    \hline
  24 & $1s^2 2s^2 2p^6 3p 3d$   & 0.0147 & $2.48\times10^{14}$  
   & 0.0219 & $5.13\times10^{13} $ \\
  23 & $1s^2 2s^2 2p^6 3s^2 3p$ & 0.318 & $6.02\times10^{13}$  
   & 0.0931 & $1.68\times10^{13} $ \\
  25 & $1s^2 2s^2 2p^6 3s$ & 0.157 & $4.50\times10^{13}$  
   & 0.305 & $8.46\times10^{13} $ \\
    \hline
  \multicolumn{2}{c}{Total} & 1 & $4.43\times10^{14}$ & 1 & 
   $2.82\times10^{14}$ \\
\hline\hline	
\end{tabular}
  \caption{Configurations with the largest contribution to the 
  radiative bound-free (bf) losses in Kr at 500 eV and 
  $10^{18}\text{ cm}^{-3}$, in the configuration average (CA) case 
  and in detailed level (DL) computation. The total population of 
  the configuration and its contribution to bf losses (erg/s/cm$^3$) 
  is given in each case.\label{tab:Kr_Ne18_dominant_contrib_bf}}
\end{table}

\begin{landscape} %
\begin{table}[htb] %
	\centering%
		\begin{tabular}{c|cccc|cccc|cccc}
		\hline\hline
 $T_e\text{ (eV)}$ & \multicolumn{4}{c|}{$<Z^*>$} & 
 \multicolumn{4}{c|}{Bound-bound losses ($10^{16}\text{ erg/s/cm}^3$)} & 
 \multicolumn{4}{c}{Bound-free losses ($10^{14}\text{ erg/s/cm}^3$)}\\[-3pt]
  (eV) & detailed &(\ref{eqn:avratgdef})&(\ref{eqn:avratgBoltzdef})&(\ref{eqn:avrat_tconfdef}) 
       & detailed &(\ref{eqn:avratgdef})&(\ref{eqn:avratgBoltzdef})&(\ref{eqn:avrat_tconfdef})
       & detailed &(\ref{eqn:avratgdef})&(\ref{eqn:avratgBoltzdef})&(\ref{eqn:avrat_tconfdef})\\
    \hline
  500 & 24.516 & 23.749 & 23.733 & 23.586 
      &  5.949 &  9.327 &  9.332 &  9.381
      &  2.823 &  4.406 &  4.431 &  4.633 \\
 1000 & 25.786 & 25.654 & 25.654 & 25.650 
      &  1.919 &  2.333 &  2.331 &  2.318 
      &  2.465 &  2.495 &  2.499 &  2.532 \\
 2000 & 28.457 & 27.826 & 27.830 & 27.860 
      &  2.553 &  3.049 &  3.053 &  3.097 
      &  3.299 &  3.095 &  3.097 &  3.110 \\
 5000 & 33.297 & 33.234 & 33.234 & 33.234 
      &  0.9903&  1.347 &  1.345 &  1.319 
      &  4.638 &  4.573 &  4.573 &  4.573 \\
\hline\hline	
\end{tabular}
  \caption{Krypton plasma properties at $10^{18}\text{ e.cm}^{-3}$. 
  The average net charge $<Z^*>$, radiative bound-bound and bound-free 
  losses are given in the detailed-level models (columns 2, 6, 10) 
  and in the configuration average approximation. This average is 
  performed using in the Boltzmann factor a configuration 
  temperature $T_c$ infinite (Eq. \ref{eqn:avratgdef}), equal to 
  the electronic temperature $T_e$ (Eq. \ref{eqn:avratgBoltzdef}) 
  or such as $T_c=0.1T_e$ (Eq. \ref{eqn:avrat_tconfdef}). 
  \label{tab:Kr_Ne18_gBoltz_vs_g}}
\end{table}
\vspace{1cm}

\begin{table}[hbt] %
	\centering%
		\begin{tabular}{c|cccc|cccc|cccc}
		\hline\hline
 $T_e\text{ (eV)}$ & \multicolumn{4}{c|}{$<Z^*>$} & 
 \multicolumn{4}{c|}{Bound-bound losses ($10^{7}\text{ erg/s/cm}^3$)} & 
 \multicolumn{4}{c}{Bound-free losses ($10^{6}\text{ erg/s/cm}^3$)}\\[-3pt]
  (eV) & detailed &(\ref{eqn:avratgdef})&(\ref{eqn:avratgBoltzdef})&
          (\ref{eqn:avrat_tconfdef}) 
       & detailed &(\ref{eqn:avratgdef})&(\ref{eqn:avratgBoltzdef})& 
          (\ref{eqn:avrat_tconfdef})
       & detailed &(\ref{eqn:avratgdef})&(\ref{eqn:avratgBoltzdef})&
          (\ref{eqn:avrat_tconfdef})\\
    \hline
 $2\times10^4$ & 65.004 & 64.541 & 64.547 & 64.571 
               &  6.370 & 1.290[6] & 1.223[6] & 6.766[5] 
               &  9.238 &  8.601 &  8.606 &  8.635 \\
 $3\times10^4$ & 67.389 & 67.263 & 67.265 & 67.254 
               &  4.866 & 6.683[6] & 6.496[6] & 4.866[6] 
               &  9.917 &  9.396 &  9.399 &  9.424 \\
\hline\hline	
\end{tabular}
  \caption{Same as table \ref{tab:Kr_Ne18_gBoltz_vs_g} but for a 
  tungsten plasma at $10^{14}\text{ e.cm}^{-3}$. The notation 1.290[6] 
  stands for $1.290\times10^6$.}
%
%
%
%
  \label{tab:W_Ne14_gBoltz_vs_g}
\end{table}
\end{landscape}

\clearpage
\listoffigures

\clearpage

\clearpage

\clearpage

\clearpage

\clearpage

\clearpage

\end{document}